\documentstyle[12pt,psfig]{article}

\begin{document}
\begin{titlepage}
\begin{flushright}\vbox{\begin{tabular}{c}
	   MRI-PHY/P9906017\\
           TIFR/TH/99-30\\
           June, 1999\\
           hep-lat/9906023\\
\end{tabular}}\end{flushright}
\begin{center}
   {\large \bf
      Screening Masses in $SU(2)$ Pure Gauge Theory
      }
\end{center}
\bigskip
\begin{center}
   {Saumen Datta\footnote{E-mail: saumen@mri.ernet.in}\\
    The Mehta Research Institute,\\
    Chhatnag Road, Jhusi, Allahabad 211019, India.\\
    and\\
    Sourendu Gupta\footnote{E-mail: sgupta@theory.tifr.res.in}\\
    Department of Theoretical Physics,\\
    Tata Institute of Fundamental Research,\\
    Homi Bhabha Road, Mumbai 400005, India.}
\end{center}
\bigskip
\begin{abstract}
We perform a systematic scaling study of screening masses in pure gauge
$SU(2)$ theory at temperatures above the phase transition temperature.
The major finite volume effect is seen to be spatial deconfinement. We
extract the screening masses in the infinite volume and zero lattice
spacing limit.  We find that these physical results can be deduced from
runs on rather coarse lattices. Dimensional reduction is clearly seen
in the spectrum.
\end{abstract}
\end{titlepage}

In a recent paper \cite{old} we examined the spectrum of screening
masses at finite temperature in four dimensional $SU(2)$ and $SU(3)$
pure gauge theories. Our primary result was that dimensional reduction
could be seen in the (gauge invariant) spectrum of the spatial transfer
matrix of the theory. In addition, we had shown that the specific
details of the spectrum precluded any attempt to understand it
perturbatively.  In this paper we present a complete set of
non-perturbative constraints on the effective dimensionally reduced
theory \cite{thomas,keijo,braaten} at a temperature ($T$) above the phase
transition temperature ($T_c$) for the $SU(2)$ case in the zero-lattice
spacing and infinite volume limit.

The study of screening masses is interesting for two reasons. First,
they are crucial to phenomenology because they determine whether the
fireball obtained in a relativistic heavy-ion collision is large enough
for thermodynamics. Second, the problem of understanding screening
masses impinges on several long-standing problems concerning the
infrared behaviour of the $T>T_c$ physics of non-Abelian gauge theories.

It is known that electric polarisations of gluons get a mass in
perturbation theory, whereas magnetic polarisations do not. Long ago,
Linde pointed out \cite{linde} that $T>0$ perturbation theory breaks
down at a finite order due to this insufficient screening of the
infrared in non-Abelian theories.  The most straightforward way to cure
this infrared divergence would be if the magnetic polarisations also
get a mass non-perturbatively. There have been recent attempts to
measure such a mass in gauge-fixed lattice computations \cite{frithjof}.

It was found long back that the solution could be more complicated and
intimately related to the dynamics of dimensionally reduced theories.
Jackiw and Templeton analysed perturbative expansions in massless and
super-renormalisable three dimensional theories \cite{jackiw} and found
that subtle non-perturbative effects screen the infrared singularities
in such theories. In a companion paper, Applequist and Pisarski
discussed the possibility that such effects might, among other things, also
give rise to magnetic masses \cite{applequist}. In fact the recent
suggestion of Arnold and Yaffe that non-perturbative terms and
logarithms of the gauge coupling may be important in an expansion of
the Debye screening mass in powers of the coupling \cite{arnold} may be
seen as an example of such non-perturbative effects. The generation of
the other screening masses are also non-perturbative. We discuss these
issues further after presenting our main results.

In this paper we report our measurements of the screening masses in the
infinite volume and zero lattice spacing limit of $SU(2)$ pure gauge
theory at temperatures of 2--4$T_c$. We found a strong finite volume
movement of one of the screening masses due to spatial deconfinement.
However, the lack of finite volume effects in the remaining channels
allowed us to extract infinite volume results from small lattices.  The
effect of a finite lattice spacing turned out to be small. We were able
to pin down all the available screening masses with an accuracy of
about 5\%.

It is necessary to set out our notation for the quantum numbers of the
screening masses. The transfer matrix in the spatial direction, $z$,
has the dihedral symmetry, $D^4_h$ of a slice of the lattice which
contains the orthogonal $x$, $y$ and $t$ directions. The irreducible
representations (irreps) are labelled by charge conjugation parity,
$C$, the 3-dimensional ($x,y,t$) parity, $P$, and the irrep labels of
$D_4$ (four one-dimensional irreps $A_{1,2}$, $B_{1,2}$ and one
two-dimensional irrep $E$). In $SU(2)$ gauge theory, only the $C=1$
irreps are realised; hence we lighten the notation by dropping
this quantum number.

Dimensional reduction implies the following pair-wise degeneracies of
screening masses---
\begin{equation}
   m(A_1^P)=m(A_2^{-P}),\qquad m(B_1^P)=m(B_2^{-P}),
   \qquad m(E^P)=m(E^{-P}).
\label{dimred}\end{equation}
After this reduction, the symmetry group becomes $C^4_v$ on the
lattice and $O(2)$ in the continuum. The latter group has two real
one-dimensional irreps--- $0_+$ and $0_-$. The first comes from the
$J_z=0$ components of even spin irreps of $O(3)$, and the second from
the $J_z=0$ components of the odd spins. There are also an infinite
number of real two dimensional irreps, ${\bf M}$, corresponding to the
$J_z=\pm M$ pair coming from any spin of $O(3)$. Dimensional reduction
associates the irreps of $D^4_h$ with those of $O(2)$ according to
\begin{eqnarray}
   m(0_+)=m(A_1^+),\;&&\; m(0_-)=m(A_1^-),\cr
   m({\bf 1})=m(E),\;&&\; m({\bf 2})=m(B_1^+)=m(B_1^-).
\label{conti}\end{eqnarray}
The final double equality is valid only when all lattice artifacts disappear.
Although $O(2)$ has an infinite tower of states, only these four masses
are measurable in a lattice simulation of the $SU(2)$
theory\footnote{There has been a first attempt to disentangle these
lattice effects and measure the higher irreps \cite{tower}.}.  Some of
the equalities in eqs.\ (\ref{dimred},\ref{conti}) may be broken by
dynamical lattice artifacts.

We studied these artifacts using ``torelon'' correlators \cite{cm}.
These are correlation functions of Polyakov loop operators in the
spatial ($P_x$ and $P_y$) and temporal ($P_t$) directions. $P_t$ and
$P_x+P_y$ transform as the scalar ($A_1^+$) of $D^4_h$, whereas
$P_x-P_y$ transforms as $B_1^+$. At zero temperature, a major part of
finite volume effects in masses can be understood (for moderate $mL$)
in terms of torelons.

The status of the $A_1^+$ torelon, $P_t$, at $T>0$ is very different
from that at $T=0$. Here, $P_t$ is the order parameter for the phase
transition, and its correlations have genuine physical meaning---
giving the static quark-antiquark potential, and hence defining the
Debye screening mass, $M_D$.  This is identical to $m(A_1^+)$ obtained
from the Wilson loop operators \cite{vold}. In this respect, the finite
temperature theory is nothing but a finite size effect.

We believe that the major part of finite volume effects in screening
masses can be understood in terms of finite temperature physics. In
simulations of $N_t\times L^2\times N_z$ lattices at a given coupling
$\beta$, when the transverse direction, $L$, is small enough, the
spatial gauge fields are deconfined. The spatial torelons $P_{x,y}$ are
order parameters for this effect. In general, large lattices, $L/N_t\gg
T/T_c$ have to be used to obtain the thermodynamic limit. Below this
limiting value of $L$, we should find strong finite volume effects, but
only in the $A_1^+$ and $B_1^+$ sectors. When such effects can be
directly measured, the $B_1^+$ loop mass is expected to be twice the
torelon mass. Whether or not similar effects are seen in the $A_1^+$
sector depends on whether the spatial torelon mass is less than
$M_D/2$. If it is, then finite volume effects should be strong,
otherwise not. We look upon torelons as convenient probes of finite
volume effects, not their cause.

We have studied screening masses for $SU(2)$ gauge theory on $N_t\times
L^2 \times N_z$ lattices with $N_z=4N_t$ at a temperature of $T=2T_c$.
We studied two series of lattices, one for $N_t=4$ and another for
$N_t=6$. For the first, we took $L=8$, 10, 12 and 16. For the second,
we chose $L=16$, 20 and 24. For $N_t=4$, a temperature of $2T_c$ is
obtained by working with $\beta=2.51$. On $N_t=6$, the choice
$\beta=2.64$ gives $T=2T_c$. The choice of lattice sizes allowed us to
investigate finite volume as well as finite lattice spacing effects at
constant physics.

We have also carried out measurements at $T=3T_c$ and $4T_c$. Since our
measurements at $2T_c$ showed that lattice spacing effects are quite
small for $N_t=4$, we restricted ourselves to this size at higher
temperatures. At $3T_c$, we worked with a $4\times24^3$ lattice. At
$4T_c$, we supplemented our earlier measurements on small
($4\times8^2\times 16$ and $4\times12^2\times16$) lattices \cite{old}
with measurements on $4\times24^3$ and $4\times32^3$ lattices.
For $N_t=4$, temperatures of $3T_c$ and $4T_c$ are attained by working
at $\beta=2.64$ and 2.74, respectively.

We used a hybrid over-relaxation algorithm for the Monte-Carlo
simulation, with 5 steps of OR followed by 1 step of a heat-bath
algorithm. The autocorrelations of plaquettes and Polyakov loops were
found to be less than two such composite sweeps; hence measurements
were taken every fifth such sweep. We took $10^4$ measurements in each
simulation except on the $6\times24^3$ lattice where we took twice as
much, and the $4\times32^3$ lattice where we took 4000.

Noise reduction involved fuzzing. The full set of loop operators
measured on some of the smaller lattices can be found in \cite{old}.
Since analyses of subsets of these operators gave identical results, we
saved CPU time on the larger lattices by measuring a smaller number of
operators.  The full matrix of cross correlations was constructed,
between all operators at all levels of fuzzing, in each irrep. A
variational procedure was used along with jack-knife estimators for the
local masses. Torelons were also subjected to a similar analysis.

\begin{table}[bhtp]\begin{center}
\begin{tabular}{|c|c|c|c|c|c|} \hline
\multicolumn{2}{|c|}{Operators} & $L=8$ & $L=10$ & $L=12$ & $L=16$ \\ \hline
Loops&$A_1^+$ & $0.71\pm0.05$ & $0.73\pm0.06$ & $0.69\pm0.04$ & $0.73\pm0.05$
     \\ \cline{2-6}
     &$A_1^-$ & $1.14\pm0.02$ & $1.01\pm0.03$ & $1.02\pm0.02$ & $0.99\pm0.02$
     \\ \cline{2-6}
     &$B_1^+$ & $1.18\pm0.03$ & $1.45\pm0.05$ & $1.62\pm0.06$ & $1.73\pm0.08$
     \\ \cline{2-6}
     &$B_1^-$ & $1.9\pm0.2$   & $1.9\pm0.1$   & $1.8\pm0.1$   & $1.8\pm0.1$
     \\ \cline{2-6}
     &$B_2^+$ & $1.8\pm0.1$   & $1.8\pm0.1$   & $1.75\pm0.05$ & $1.76\pm0.09$
     \\ \cline{2-6}
     &$B_2^-$ & $1.9\pm0.2$   & $1.9\pm0.2$   & $1.8\pm0.2$   & $1.9\pm0.2$
     \\ \cline{2-6}
     &$E^+$   &               &               & $1.80\pm0.07$ & $1.8\pm0.1$
     \\ \cline{2-6}
     &$E^-$   &               &               & $1.68\pm0.07$ & $1.7\pm0.2$
     \\ \hline
Torelon&$P_t$ & $0.76\pm0.04$ & $0.75\pm0.05$ & $0.80\pm0.02$ & $0.73\pm0.08$ 
     \\ \cline{2-6}
       &$P_x+P_y$ & $0.46\pm0.02$ & $0.7\pm0.1$ & $0.9\pm0.1$ & - 
     \\ \cline{2-6}
       &$P_x-P_y$ & $0.48\pm0.02$ & $0.7\pm0.1$ & $0.8\pm0.1$ & - \\
 \hline
\end{tabular}\end{center}
\caption[dummy]{Values of $m$ on $4\times L^2\times16$ lattices in units of
   inverse lattice spacing, $1/a=4T$ at $T=2T_c$. 
   Local masses could be followed to distance
   $z\approx4/m$. Blanks in the table mean that the operators were not
   measured, and an entry of ``-'' denotes that the measurements
   were too noisy to yield a screening mass.}
\label{res.tbl4}\end{table}

Our measurements at $2 T_c$ for $N_t=4$ are reported in 
Table \ref{res.tbl4}.  We
can measure torelons for fairly large values of $L/N_t$.  Twice the
$A_1^+$ spatial torelon screening mass is greater than that obtained from
$P_t$. Hence $m(A_1^+)$ obtained from loops is equal to the latter and
therefore shows no finite volume effect.  The $A_1^+$ and $B_1^+$
spatial torelons have equal screening masses. The $B_1^+$ loop
screening mass closely equals twice the $B_1^+$ torelon mass, and hence
shows a systematic dependence on $L$. Finite volume effects are absent
in all the other channels, as expected. For the $L=16$ lattice for
$N_t=4$, the torelon is not measurable, and finite volume effects are
under control. At this largest volume dimensional reduction and
continuum physics are visible since the equalities in
eqs.\ (\ref{dimred},\ref{conti}) are satisfied.

We have investigated finite lattice spacing effects by making the same
measurements at the same physical temperature on lattices with
$N_t=6$.  The measurement of $m(E^-)$ turns out to be rather noisy.
Since we had observed on the coarser lattice that $m(E^+)=m(E^-)$, we
saved on CPU time by dropping the measurement of the $E^-$ screening
mass on the $N_t=6$ lattices.  Our results on the finer lattice are
collected in Table \ref{res.tbl6}.  Again, dimensional reduction and
continuum physics is visible because the equalities in
eqs.\ (\ref{dimred},\ref{conti}) are satisfied on the largest lattice.

\begin{table}[htbp]\begin{center}
\begin{tabular}{|c|c|c|c|c|} \hline
\multicolumn{2}{|c|}{Operators} & $L=16$ & $L=20$ & $L=24$ \\ \hline
Loops&$A_1^+$ & $0.46\pm0.02$ & $0.51\pm0.01$ & $0.51\pm0.02$ \\ \cline{2-5}
     &$A_1^-$ & $0.69\pm0.01$ & $0.68\pm0.02$ & $0.68\pm0.02$ \\ \cline{2-5}
     &$B_1^+$ & $0.98\pm0.03$ & $1.07\pm0.06$ & $1.08\pm0.09$ \\ \cline{2-5}
     &$B_1^-$ & $1.23\pm0.06$ & $1.17\pm0.07$ & $1.19\pm0.08$ \\ \cline{2-5}
     &$B_2^+$ & $1.23\pm0.08$ & $1.23\pm0.09$ & $1.22\pm0.06$ \\ \cline{2-5}
     &$B_2^-$ & $1.28\pm0.09$ & $1.27\pm0.05$ & $1.25\pm0.11$ \\ \cline{2-5}
     &$E^+$   & $1.25\pm0.03$ & $1.3 \pm0.1 $ & $1.24\pm0.08$  \\ \hline
Torelon&$P_t$ & $0.46\pm0.02$ & $0.52\pm0.02$ & $0.51\pm0.02$ \\ \cline{2-5}
       &$P_x+P_y$ &$0.32\pm0.17$ & - & - \\ \cline{2-5}
       &$P_x-P_y$ &$0.37\pm0.16$ & - & - \\ \hline
\end{tabular}\end{center}
\caption[dummy]{Values of $m$ on $6\times L^2\times24$ lattices in units
   of the inverse lattice spacing $1/a=6T$ at $T=2T_c$.
   Local masses could be followed to distance
   $z\approx4/m$. An entry of ``-'' in the table denotes that the 
   measurements were too noisy to yield a screening mass.}
\label{res.tbl6}\end{table}

From the data collected in Tables \ref{res.tbl4} and \ref{res.tbl6} it
is clear that the physical ratio $m/T$ is the same with both lattice
spacings, for loop masses. Hence finite lattice spacing effects are
under control.  This result is consistent with zero temperature lattice
measurements which show that at these lattice spacings, ratios of
physical quantities are independent of the spacing.

\begin{figure}[hbt]
\begin{center}\leavevmode
\psfig{figure=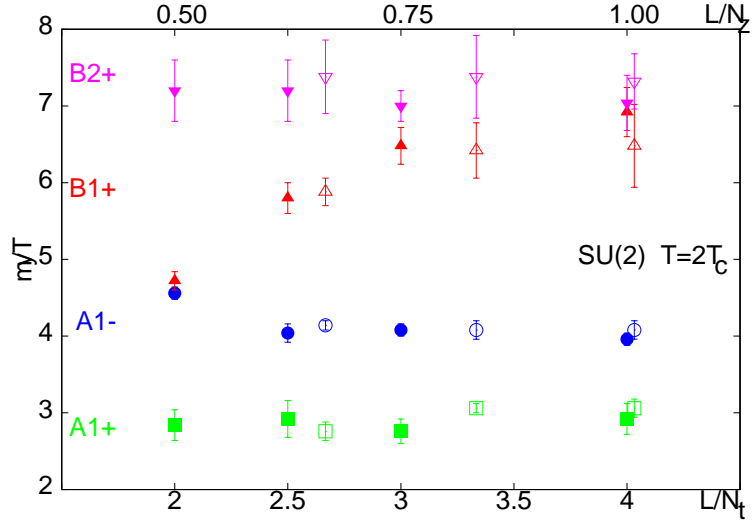,height=7cm,width=10cm}
\end{center}
\caption[dummy]{Values of $m/T$ using different operators as a function of
   $L/N_t$ (or $L/N_z$), the scaled transverse spatial size of the lattice.
   The open symbols are for $N_t=6$ lattices, filled symbols for
   $N_t=4$. In order to improve visibility, the data at $L/N_t=4$ for
   $N_t=6$ has been displaced slightly to the right.}
\label{fmass}\end{figure}

In Figure \ref{fmass} we illustrate the nature of the finite volume
effects.  The lack of movement in $m(A_1^+)$, $m(A_1^-)$ and $m(B_2^+)$
is obvious. We have used the fact of dimensional reduction to prune the
amount of data that has to be displayed in the graph. Note that the
data show that $m({\bf 2})$ can be estimated by measuring any of the
$B$ irrep screening masses, apart from the $B_1^+$, at small volumes.
Note also that $m(B_1^+)/T$ scales with either $L/N_t$ or $L/N_z$.
However, for $N_t=6$, it becomes difficult to measure the torelon
correlator at fairly small value of $L/N_t$. This supports our earlier
statement that the torelon is a measure, not the cause, of finite
volume effects.

In the $SU(3)$ pure gauge theory, which has a first order phase
transition, the simple equalities $N_z/N_t,L/N_t>T/T_c$ are sufficient
to remove finite volume effects \cite{old}.  The observed slow finite
volume movement of $m(B_1^+)$ is special to the $SU(2)$ gauge theory,
which has a second order finite temperature phase transition.  As a
result, the above constraints on the lattice sizes are compounded by
two separate systematics of second order phase transitions. The first
is that there are precursor effects which cause masses to decrease at
temperatures less than $T_c$; the second that part of this decrease is
power-law singular in $N_z$ at fixed $\beta$. Consequently, in the
$SU(2)$ theory we can at best state the more stringent conditions
$L/N_t=N_z/N_t\gg T/T_c$.

\begin{figure}[htb]
\begin{center}\leavevmode
\psfig{figure=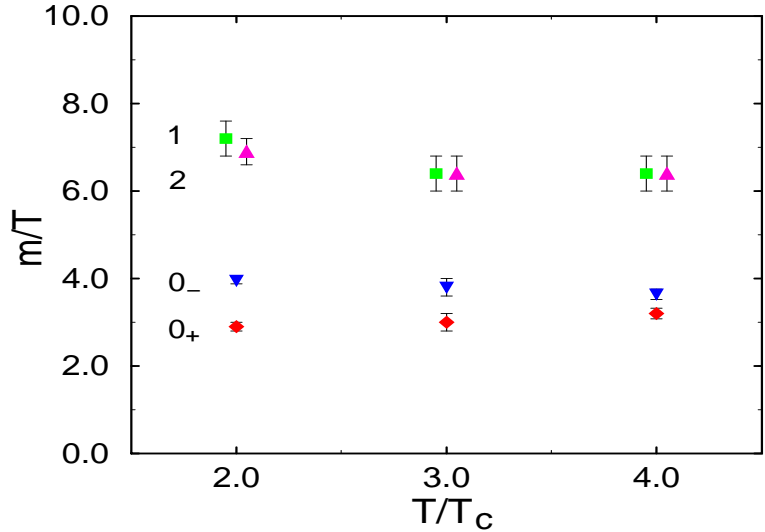,height=7cm,width=10cm}
\end{center}
\caption[dummy]{Values of $m/T$ for $2T_c\le T\le 4T_c$, showing the 
   near-constancy of the ratios in this temperature range. In order to 
   improve visibility, the points for $m({\bf 1})/T$ have been shifted
   slightly to the left and those for $m({\bf 2})/T$ slightly to the 
   right.}
\label{fg.massflow}\end{figure}

In our measurements with $N_t\times N_s^3$ lattices at higher
temperatures, we found that the lattice artifact in $m(B_1^+)$ persists
for the largest values of $N_s/N_t$ that we had. At both the higher
temperatures no other finite volume effects were seen within the
precision of the measurements. As a result, we were able to estimate
all the four screening masses listed in eq.\ (\ref{conti}). The ratios
$m/T$ are seen to be approximately constant in this temperature range.
This is illustrated in Fig. \ref{fg.massflow}.

\begin{figure}[hbt]
\begin{center}\leavevmode
\psfig{figure=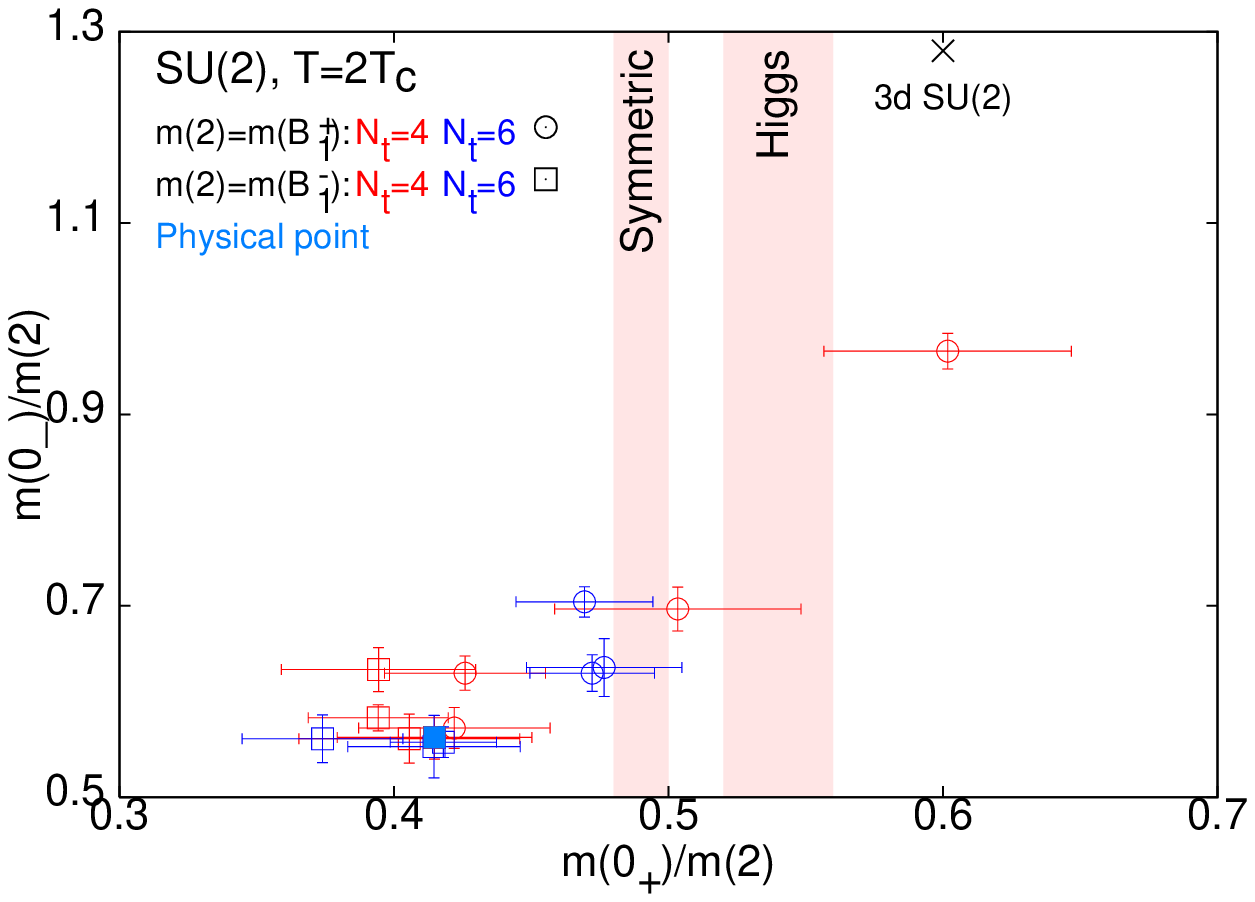,height=7cm,width=10cm}
\end{center}
\caption[dummy]{Finite size movement of ratios of screening masses with two
   different identifications of the screening mass $m({\bf 2})$. The physical
   point shown should be understood to have error bars compatible with those
   for the other points. The point for the 3-d pure gauge theory \cite{teper}
   is denoted by the cross. The vertical bands show the measured $1-\sigma$
   ranges for the mass ratio $m(0_+)/m(\bf2)$ in an $SU(2)$ scalar-gauge
   theory \cite{owe} in the symmetric and Higgs phases.}
\label{frat}\end{figure}

The four masses that we have extracted from simulations of the 4-d
theory represent the maximum information available non-perturbatively
to constrain the effective 3-d theory. We found it instructive to
display the same data in Figure \ref{frat} as a plot of the ratio
$m(0_+)/m({\bf 2})$ against $m(0_-)/m({\bf 2})$.  The finite volume
movement in these numbers is fairly large if the denominators are
estimated through $m(B_1^+)$. However, as shown, the movement is much
reduced if $m(B_1^-)$ is used as an estimator of $m({\bf 2})$.  Since
the continuum and thermodynamic limit is pretty well pinned down, the
figure also serves well to compare the 4-d theory with different 3-d
theories.

The point for the 3-d $SU(2)$ pure gauge theory in the infinite
volume and for zero lattice spacing \cite{teper} is shown in the figure.
It is clear that this is not the appropriate
effective theory. This result is expected, since a
perturbative mode counting shows that the effective three dimensional
theory must contain a gauge field and a scalar field that transforms
adjointly under gauge transformations, and the scalar field does not
decouple completely from the theory even at high temperatures
\cite{landsman}.

The vertical bands in Figure \ref{frat} come from measurements in a 3-d
$SU(2)$ gauge theory with a fundamental scalar in both the symmetric and
Higgs phases of this theory \cite{owe}.  It is not surprising that the
ratio $m(0_+)/m(\bf2)$ in either phase of this theory does not agree
with our measurements at $2T_c$, in view of the arguments already
presented.

In \cite{keijo} a super-renormalisable 3-d theory of $SU(2)$
gauge fields and an adjoint scalar, with three couplings, was suggested
as the effective theory. Matching two of these couplings in a
perturbation expansion, the screening masses were computed through a
simulation of the 3-d theory. It turned out that at couplings
corresponding to a temperature of $2T_c$, $m({\bf 1})/m(0_+)=1.6\pm0.2$
as opposed to the value $2.4\pm0.2$ that we measure. Whether better
agreement can be obtained by fine-tuning the third coupling remains as
a future exercise.  If three couplings can be tuned to reproduce four
masses, then this would vindicate the perturbative approach to matching
espoused in \cite{keijo}.

However, until such a demonstration is made, there are questions whether
this procedure is viable at
$T\approx2T_c$. A direct measurement suggests that the gauge coupling
is larger than unity, $g^2/2\pi \approx0.53$, even for $T=2T_c$
\cite{bali}. A related statement has been made based on a recent study
of the Debye mass--- that higher orders in the perturbative series
become numrically smaller only at $T\approx10^7T_c$ \cite{owe2}. A similar
statement comes from attempts to find the region of validity of
the perturbative expansion of the free
energy in a non-Abelian plasma \cite{free}, which give $T>10^5T_c$.  It
has recently been suggested \cite{braaten2} that effects associated
with screening and damping should be resummed to all orders in $g$,
if perturbation theory is to behave reasonably at
$T\approx2T_c$.  We have earlier concluded that the screening masses we
observe cannot be obtained perturbatively \cite{old}.  The fact that
our measurements show $m/T>2\pi$ in some channels also indicates that
the perturbative matching procedure may not be useful, since
dimensional reduction works only if modes with energy $2\pi T$ or more
decouple \cite{thomas,keijo}.

There are alternatives to perturbation theory. One interesting method
would be to use gauge invariant composites directly to construct the
effective theory. Phenomenology of this kind was used long ago to
examine lattice data on the energy density for $T>T_c$ $SU(3)$ gauge
theory \cite{singlet}. A more sophisticated attempt of this kind was
tried in \cite{ashok}, but needed the machinery of large-N theories to
control the expansion.

The question of the compositeness of screening masses is closely
related. Note that the 3-d adjoint Higgs, $A_t$, is in the $0_-$ irrep
of $O(2)$, and the 3-d gauge field, ${\bf A}$, in the ${\bf 1}$. The
gauge invariant $0_+$ can be seen in correlations of the composite
operators ${\cal O}_1={\rm Tr}(A_t^2)$ and ${\cal O}_2={\rm Tr}({\bf
A}\cdot{\bf A})$, as well as higher dimensional operators.  The gauge
invariant $0_-$ screening mass can be seen, for example, in
correlations of ${\cal O}_3={\rm Tr}(A_t^3)$. The composite operators
corresponding to the remaining gauge invariant screening masses can
also be easily written down.  Nadkarni had shown by explicit
computation that ${\cal O}_1$ and ${\cal O}_2$ mix at order $g^4$,
where $g$ is the 4-d gauge coupling \cite{nadkarni}.  Hence, the
characterisation of $m(0_+)$ as being due to electric phenomena is a
perturbative statement, and is more or less correct according to how
large $g$ is at the temperature that concerns us. Similar
problems occur in the other channels as well.

In summary, we identified the only source of large finite volume
effects in the determination of screening masses at $T>T_c$ in $SU(2)$
pure gauge theory. These are due to spatial deconfinement and can be
conveniently
studied using torelons. Finite lattice spacing effects turn out to be
easy to control. We found that rather small and coarse lattices can be
used to obtain a good measurement of the physical screening masses,
provided one ignores the $B_1^+$ channel. Our best estimates are
shown in Figure \ref{fg.massflow}.  Dimensional reduction, as expressed
non-perturbatively in eqs.\ (\ref{dimred},\ref{conti}), is seen in the
temperature range 2--4$T_c$.

We would like to thank Owe Philipsen for a discussion.

\vfill\eject

\end{document}